\documentclass[aps,prd,twocolumn,floatfix,altaffilletter,superscriptaddress,preprintnumbers,
tightenlines,showpacs,showkeys,nofootinbib,notoccite]{revtex4-1}
\usepackage[utf8]{inputenc}
\usepackage[colorlinks=true,citecolor=blue,linkcolor=blue]{hyperref}
\usepackage[normalem]{ulem}
\usepackage{amsmath,amssymb, mathrsfs,esvect}
\usepackage{epsfig}
\usepackage{graphicx}               
\usepackage{url}
\usepackage{color}
\usepackage{slashed}
\usepackage{multirow}
\usepackage{placeins}
\usepackage[dvipsnames]{xcolor}
\usepackage{epstopdf}
\usepackage{soul}
\usepackage{tikz}
\usepackage[capitalise, english]{cleveref}
\usepackage{siunitx}
\usepackage{xspace}
\usepackage{booktabs}
\usepackage[capitalise]{cleveref}
\usetikzlibrary{trees}
\usetikzlibrary{decorations.pathmorphing}
\usetikzlibrary{decorations.markings}

\newcommand\myshade{80}
\colorlet{mylinkcolor}{ForestGreen}
\colorlet{mycitecolor}{Red}
\colorlet{myurlcolor}{violet}

\hypersetup{
	linkcolor  = mylinkcolor!\myshade!black,
	citecolor  = mycitecolor!\myshade!black,
	urlcolor   = myurlcolor!\myshade!black,
	colorlinks = true
}

\definecolor{jblue}{RGB}{20,50,100}
\definecolor{npurple}{RGB} {153, 51, 204}
\definecolor{wred}{RGB}{217,0,56}
\definecolor{white}{RGB}{255,255,255}
\definecolor{forestgreen}{HTML}{228B22}

\allowdisplaybreaks
\usepackage{tikz,xcolor,hyperref}

\definecolor{lime}{HTML}{A6CE39}
\DeclareRobustCommand{\orcidicon}{\hspace{-1mm}
	\begin{tikzpicture}
		\draw[lime, fill=lime] (0,0) 
		circle [radius=0.16] 
		node[white] {{\fontfamily{qag}\selectfont \tiny \,ID}};
		\draw[white, fill=white] (-0.0525,0.095) 
		circle [radius=0.007];
	\end{tikzpicture}
	\hspace{-3mm}
}

\foreach \x in {A, ..., Z}{\expandafter\xdef\csname orcid\x\endcsname{\noexpand\href{https://orcid.org/\csname orcidauthor\x\endcsname}
		{\noexpand\orcidicon}}
}

\usepackage{tikz}
\usepackage{tikz-feynman}
\tikzfeynmanset{compat=1.0.0}

\begin{document}
	
\title{Freeze-in dark matter in neutron stars}
	
	\author{Maxim Pospelov}
	\email{pospelov@umn.edu}
		\affiliation{William I. Fine Theoretical Physics Institute, School of Physics and Astronomy\\ University of Minnesota, Minneapolis, MN 55455, USA}
	\affiliation{School of Physics and Astronomy, University of Minnesota, Minneapolis, MN 55455, USA}
	\author{Samya Roychowdhury}
	\email{roych027@umn.edu}
	\affiliation{School of Physics and Astronomy, University of Minnesota, Minneapolis, MN 55455, USA}
\date{\today}
	
\begin{abstract}
Every neutron star is born in the process of core-collapse supernova explosion that, for a brief moment, reproduces conditions of the early Universe with temperatures $T\sim O(30\rm\,MeV)$. We calculate the production of Dark Matter $\chi$ from the SM particles in such events, SM\,$\to\chi\bar\chi$, for the freeze-in range of couplings, $\alpha_{\rm FI} \sim O(10^{-26}) $,  finding that $O(10^{-6})$ $\chi$'s per nucleon is produced. The strong gravitational potential well of the neutron star retains a substantial fraction of these particles that will eventually undergo the reverse process of energy injection, $\chi\bar\chi\to$\,SM. This may lead to the abnormal energy injection creating observable signatures such as late-time  heating of the neutron stars. To demonstrate the power of this method, we construct a set of simple dark matter models coupled to lepton currents, and show that neutron stars provide unique constraints on parameter space that otherwise cannot be accessed by other means, probing effectively the scattering cross sections with the SM in the ballpark of $\sigma_{\chi\,\rm SM} \propto O(10^{-70})\,\rm cm^2$.  
\end{abstract}
\maketitle

{\bf Introduction.}
The Standard Model (SM) of particles and fields, despite its remarkable success, needs to be augmented to describe the dark matter (DM) in the Universe, that manifests itself in a variety of observations consisting of DM exerting a gravitational pull on the SM-built objects. There are plenty of theoretical ideas about the origin of DM \cite{Cirelli:2024ssz}, and in this paper we will adhere to one of the simplest possibilities, the so-called {\em freeze-in} (FI) dark matter pointed out in a number of early works (see {\em e.g.} \cite{Dodelson:1993je,McDonald:2001vt}) and generalized in Ref.\,\cite{Hall:2009bx}.

Assuming that the Universe initially had a negligible amount of DM, the thermal processes during the Hubble expansion in the very early epoch can generate a small population of DM particles which with time will gain in energy over SM radiation, and will come to dominate the Universe. The process of freeze-in is slow/inefficient ``by design" due to very tiny couplings, which keep the rate to populate DM particles below the Hubble rate at all times. This is one of the hardest classes of DM to probe using non-gravitational interactions due to the smallness of couplings. However, a few exceptions are known: for example, sterile neutrinos in the keV range with a small mixing angle with active species can be probed and indeed excluded \cite{Abazajian:2017tcc} in the minimal scenario \cite{Dodelson:1993je} using the X-ray signature from their late decays. Among the freeze-in models where stable DM particles are pair produced through SM\,$\to\chi\bar\chi$, scenarios with 
very light/massless vector mediators $A'$ coupled to electromagnetic currents (also known as ``dark photons") can be probed in the most sensitive dark matter direct detection experiments \cite{DAMIC-M:2025luv,PandaX-4T:2021bab,LZ:2022lsv,XENON:2023cxc}. This relies on the enhancement of the scattering cross section by powers of an inverse $m_{A'}$ (or powers of a scattering momentum transfer). If, on the other hand, mediator and DM have similar mass, $m_\chi \sim m_{A'}$, the scattering cross sections per nucleon or per electron are generally much below the $10^{-50}\,{\rm cm}^2$ benchmark, and cannot be probed by the ongoing/future direct detection efforts. 

In this paper, we argue that the neutron stars (NS) offer a unique window on some class of freeze-in DM models that cannot be probed in any other way. NS are born in the core-collapse supernova (SN) events which can be viewed as the most direct ``simulation" of the hot Big Bang conditions in the late Universe. Exotic particles produced in SN can be trapped by the gravity of newborn NS, and if stable, $\chi/\bar\chi$ population can be preserved over the astronomical time scales. Inevitable cooling of NS forces this population to undergo the reverse process, $\chi\bar\chi\to$\,SM, giving rise to particle emission signatures, and late heating of NS. Naturally, the most relevant mass range that NS can probe this way lies between MeV to 100 MeV scales, {\em i.e.} commensurate to star temperature during the SN explosion.  Earlier, related ideas were explored for the population of unstable beyond-SM particles in the SN/NS \cite{Hannestad:2001jv,Hannestad:2003yd}. Similar mechanisms were also applied for solar emission, where the gravitational pull is much weaker, and the thermal range points to keV scale particles as a reasonable target \cite{VanTilburg:2020jvl,Lasenby:2020goo,Giovanetti:2024rxi}. Our paper initiates studies of the freeze-in DM in SN/NS as perhaps the most motivated physics case of DM models that so far remain unconstrained. 

The late time heating of NS by DM captured from galactic halo has been an intense topic of studies \cite{Bramante:2023djs}, even though the prospects of probing freeze-out DM this way are for the distant future. The main reason is that the galactic density of DM is very low, and even for the 100\% trapping efficiency and annihilation, the amount of heat supplied cannot change in any appreciable way the temperature of the coldest known NS/pulsar \cite{Guillot:2019ugf}, PSR J2144-3933, which currently is limited to be below $kT\simeq 3\,$eV. Despite the extreme smallness of the FI couplings, the amount of FI DM produced in the SN explosion, and associated DM energy stored in the gravitational well of NS, vastly exceeds the amounts that can be captured from the outside. This offers a pathway for putting constraints on FI dark matter using {\em existing} NS observations.

\begin{figure}[t]
\includegraphics[width=0.36\textwidth]{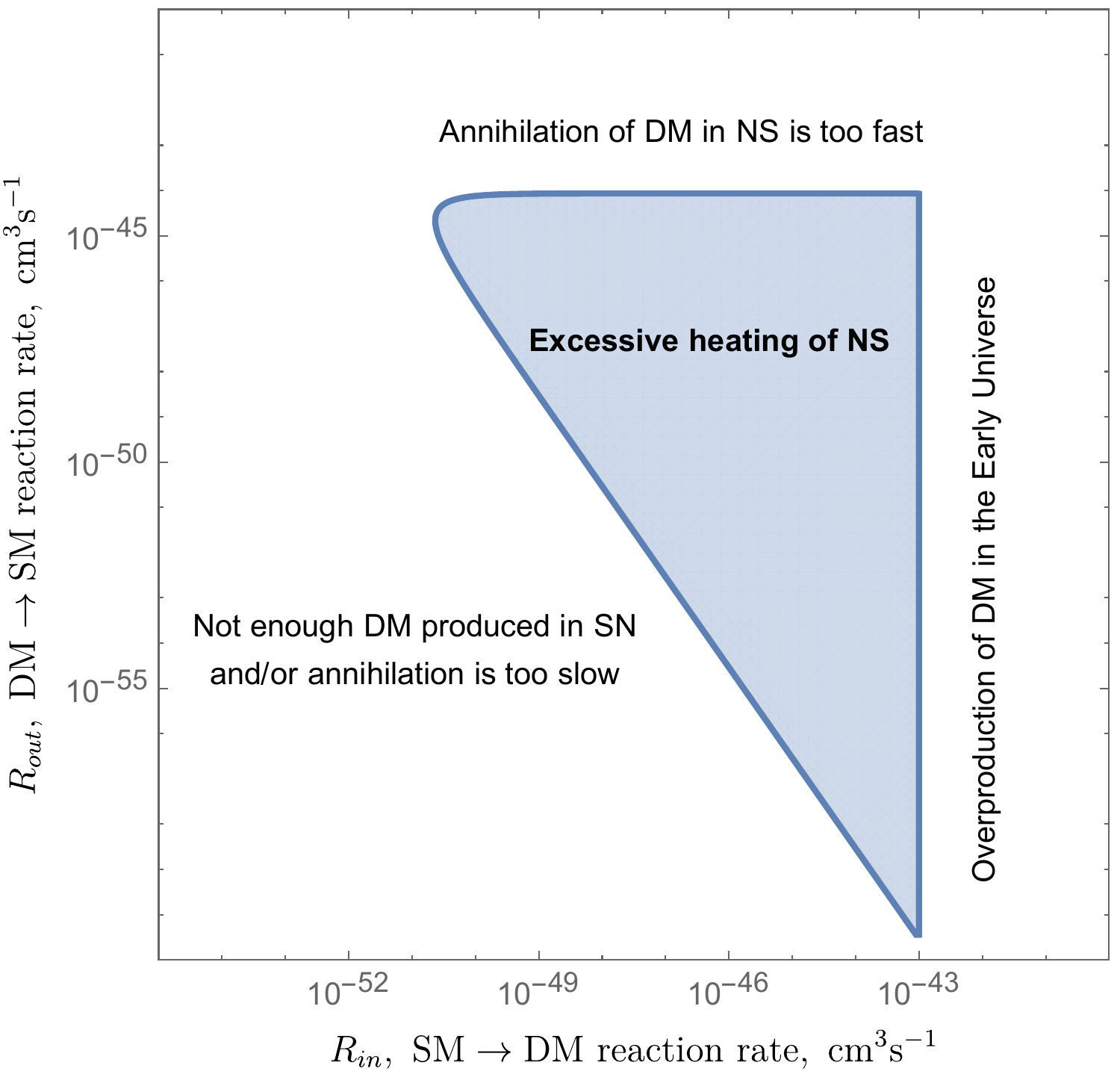}
\caption{Schematic diagram showing in- and out- reaction rates that can be relevant for the FI DM production, and for the late-time heating of NS. }
\label{fig:SM<->DM}
\end{figure}	

To provide a schematic picture of sensitivity to DM in the right mass range ({\em e.g.}$\sim 30\,\rm MeV$), we address a maximally simplified parameter space, in terms of the annihilation rates of SM particles into DM, and the reverse. Calling some generalized reaction rates, $R_{in}= \langle \sigma_{\rm SM\to \chi\bar\chi} v \rangle $ and $R_{out}= \langle \sigma_{\chi\bar\chi\to \rm SM } v \rangle $, we can adopt very crude estimates for the fractional amount of DM generated, and the amount of heat generated per volume per time in the reverse process, 
\begin{eqnarray}
\frac{n_{\chi,t\simeq 0}}{n_{\rm SM}}\propto R_{in} n_{\rm SM} \times \Delta t;~~\frac{dQ}{dVdt} \propto m_\chi \times R_{out} n_{\chi}(t)^2,
\label{estimates}
\end{eqnarray}
where $n_\chi$ and $n_{\rm SM}$ stand for number densities of DM and SM particles respectively.
In these estimates, it is assumed that the energy density of DM is dominated by its rest mass, and that the initial abundance of DM $n_{\chi,t\simeq 0}$ generated at $T\sim m_\chi $ is much smaller than a typical SM abundance $n_{\rm SM}$. The interval of time available for producing DM is different in the stars and in the early Universe. In stars, it will be on the order of the maximally hot stage of explosion, $\Delta t \sim t_{expl} \sim O(20\,\rm s)$, and in the early Universe the longest $\Delta t$ will be associated with the Hubble time $t_{\rm H}= H(T)^{-1}$, at $T\sim m_\chi$. For 30\,MeV dark matter, this Hubble time is about three orders of magnitude shorter than $t_{expl}$. Assuming that most of the DM is retained in the gravitational field of the NS and its depletion occurs via self-annihilation, one can solve for $n_{\chi}(t)$. For $R_{out} \to 0$, $n_{\chi}(t)\simeq n_{\chi,t\simeq 0}$, while in the opposite limit, 
the amount of dark matter that survives at the time $t$ is found by solving  $dn_{\chi}/dt = - R_{out}n_\chi^2 $. 

Taking $n_{\rm SM}$ in the early Universe to be $\propto T^3$, and inside the NS as $10^{39}\,{\rm cm}^{-3}$, we evaluate $R_{in}$ and $R_{out}$ leading to the freeze-in abundance of DM, and to appreciable heating of NS in excess of Ref. \cite{Guillot:2019ugf} limit at $t_{star} \sim 10^{16}$\,s.  Notwithstanding the crudeness of the above description (which assumes 100\% efficiency of gravitational capture, and 100\% heat production),  Fig.\ref{fig:SM<->DM} shows that there is a range of parameters where late time heating of NS could constrain FI models. The right face of the triangular region of interest corresponds to $R_{in}$ that would roughly saturate the DM abundance. Along that line, $R_{out}\ll R_{in} $, so that only models where the reverse annihilation is delayed can be probed. It is also easy to see that the relevant rates that could be probed via this method are $\sim$20 orders of magnitude below typical WIMP (weakly interacting massive particle) rates, and there are no alternative ways to constrain FI models. 

In what follows we formulate a very specific example of FI DM model that can be probed through the {\em existing} NS data, and where the FI cross section is much stronger than the reverse. 

{\bf FI model with leptonic force mediation.} In order to illustrate our case in more detail, we build a model with stable DM, and the cosmological abundance achieved via the freeze-in. To maximally simplify the description of processes in NS, we chose a model based on additional gauge group acting on the second and third generation of leptons. The full Lagrangian contains three structural pieces: gauged dark sector, gauged $L_\mu-L_\tau$ and the portal interaction between the two sectors. 
\begin{eqnarray}
{\cal L} &=&  {\cal L}_{L_\mu-L_\tau} + {\cal L}_{dark} + {\cal L}_{portal} ,\\
{\cal L}_{L_\mu-L_\tau} &=& \sum_{j=\mu,\tau}\bar L_{(j)}\gamma_\alpha (i \partial_\alpha \pm g_L X_\alpha) L_{(j)} 
\\
 {\cal L}_{dark} &=& \bar \chi (\gamma_\alpha (i \partial_\alpha + g_d V_\alpha)-m_\chi)\chi \\
 {\cal L}_{portal}&=& \varepsilon \frac12 X_{\mu\nu}V_{\mu\nu}
\end{eqnarray}
The sign $\pm$ is opposite between the $\mu$ and $\tau$ flavor, which leads to anomaly cancellations. 
Altogether, we have several new states: DM particles $\chi$ and $\bar\chi$, and two vector mediators $X_\mu$ and $V_\mu$, and the entire parameter space is given by 
$\{m_\chi,m_V,m_X,g_L,g_d,\varepsilon\}$. All masses are assumed to be in the MeV to 100's of MeV window, and all coupling constants are very small, and therefore not challenged by direct probes.  Moreover, in this paper, we will assume that $m_X \ll m_V $ and $m_V = 3 m_\chi$. The latter choice is quite common throughout ``dark sector" literature, although of course not crucial \cite{Berlin:2020uwy}. 

\begin{figure}[t]
\includegraphics[width=0.36\textwidth]{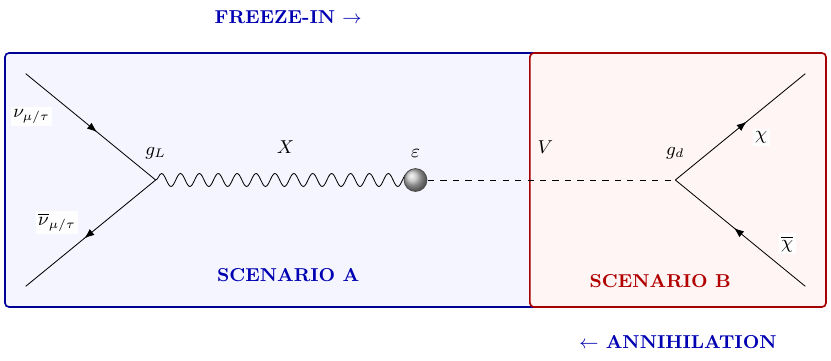}
\caption{Main Feynman diagram responsible for the production of DM pairs. The reverse process leads to the annihilation. The freeze-in sub-processes that regulate the rate of production are colored differently. For scenario A, $\alpha_{\rm FI} \propto g_L^2 \varepsilon^2$, and for B, $\alpha_{\rm FI} \propto g_d^2 $. The reverse annihilation of cooled $\chi\bar \chi$ pairs occurs via an off-shell mediator, and has $\sigma_{ann}\propto  g_d^2 g_L^2 \varepsilon^2$ scaling.}
\label{fig:F_diagrams}
\end{figure}

The main thrust of the model is the on-shell production of $V$, which decays to $\chi\bar\chi$ pairs and populates DM states on very short times scales. 
The diagram responsible for this process is shown in Fig.\,\ref{fig:F_diagrams}. Depending on the respective size of the coupling constants, we would like to point out two regimes for the freeze-in scenarios: \\
{\em Scenario A:} In this case, the $\varepsilon$ parameter is tiny, and the main freeze-in process is $2\to 1$, $\nu\bar\nu \to V$ followed by the fast decay to $\chi\bar\chi$. \\
{\em Scenario B:} In this case, the $\varepsilon$ parameter is small but sufficient for full thermalization of $V$, and the main freeze-in process is $1\to 2$, by the  out-of-equilibrium decay $V\to\chi\bar\chi$ due to a tiny $g_d$.
Both scenarios assume that $X$ vector particles are thermalized. 

The calculation of FI production of $\chi$ particles is relatively straightforward. In scenario B, it consists of calculating the rate of $V$ decay integrated over thermal history with an appropriate weight factor. In cosmological setting, the corresponding calculations for a massive dark photon production are well-known \cite{Redondo:2008ec,Fradette:2014sza}. Here, we adopt these calculations for the production of $V$. It turns out that both scenarios lead to very closely matched production rates if expressed in terms of effective FI coupling defined as: 
\begin{eqnarray}
    A:~ \alpha_{\rm FI} \equiv \frac{g_L^2\varepsilon^2}{4\pi}\times \frac{m_V^4}{(m_V^2-m_X^2)^2};~~B: \alpha_{\rm FI} \equiv \frac{g_d^2}{4\pi}.
\end{eqnarray}
The production mechanism in scenario A is driven by fusion of muon and tau neutrinos (above 210\,MeV one may also include charged muons). In the approximation of Ref. \cite{Fradette:2014sza}, the final abundance is given by 
\begin{eqnarray}
Y_\chi + Y_{\bar \chi} =  \frac{n_\chi + n_{\bar\chi}}{s} &=& \frac{\alpha_{\rm FI}}{\pi^2} m_V^3 \int \frac{dT}{s H}K_1(m_V/T)~~~ \nonumber\\ 
 &=& \frac{\alpha_{\rm FI} M_{\rm Pl}}{m_\chi} \times 6\times 10^{-3}
\end{eqnarray}
where $s$, $H$  are the entropy density and the Hubble rate, that gives the appearance to Planck mass in this formula, $M_{\rm Pl} =G_N^{-1/2}$. Comparing this to the observed DM abundance and assuming that $\chi$ particles saturate it, 
\begin{eqnarray}
    Y_\chi + Y_{\bar \chi} =4\times 10^{-8} \times\frac{10\,\rm MeV}{m_\chi} ~\to~ \alpha_{\rm FI} \simeq 6\times 10^{-27}. 
    \label{a_FI_cosmo}
\end{eqnarray}
Much above the dimuon threshold in $m_V$, this optimal value would have to be decreased by less than a factor of 2. Repeating a very similar calculation for scenario B (thermalized $V$ decaying to DM pairs with a sub-Hubble rate), we find the result that is numerically very close to (\ref{a_FI_cosmo}). While approximately similar values for $\alpha_{\rm FI}$ are expected, a close equality is an artifact of $m_V=3m_\chi$ choice.

Taking (\ref{a_FI_cosmo}) as guidance for the cosmologically relevant value of the freeze-in coupling, we proceed to calculate the emission of DM particles in the SN explosion. We take the examples of numerical simulations of the explosion process, such as Refs. \cite{Roberts:2012zza,Sumiyoshi:2022uoj}. Due to muon and tau neutrinos being the main source of exotic particles, the main observable that regulates the production rate is $T(x,t)$, temperature depending on time $t$ and some form of the radial coordinate, {\em e.g.} the enclosed mass $x = M(r)/M_{\rm NS}$. During the explosion, we neglect the subdominant muon population, and small but non-vanishing chemical potential for $\nu_\mu$. Calculation closely mirror the early Universe production, and the total number of particles {\em per baryon} produced this way can be found by evaluating this integral:
\begin{eqnarray}
\label{simplified}
    \frac{N_\chi + N_{\bar \chi} }{N_{B,\rm NS}} \simeq  
2.5\times 10^{-6}\times \frac{\alpha_{\rm FI}}{6\times 10^{-27}} \times \left(\frac{m_V}{30\,\rm MeV}\right)^3\\\nonumber\times\int\frac{dt}{10\,\rm s}\int_0^1dx \,\frac{(100\,\rm MeV)^3}{n_B(x,t)}\times\frac{T(x,t)}{10\,\rm MeV} \times K_1\left(\frac{m_V}{T(x,t)}\right).
\end{eqnarray}
In this expression, $x$ is the fraction of enclosed mass coordinate, $n_B(x,t)$ is the space and time dependent baryon density of the proto-NS, that is also provided in numerical simulations \cite{Roberts:2012zza}. Famously, the temperature during the explosion reaches maxima of $\sim 40$\,MeV lasting for $\sim 10$-$20$\,s. Expression (\ref{simplified}) shows that in the cosmologically relevant range of $\alpha_{\rm FI}$, the total emission of FI DM is small and cannot change the energetics of the explosion. 

\begin{figure}[t]
\includegraphics[width=0.42\textwidth]{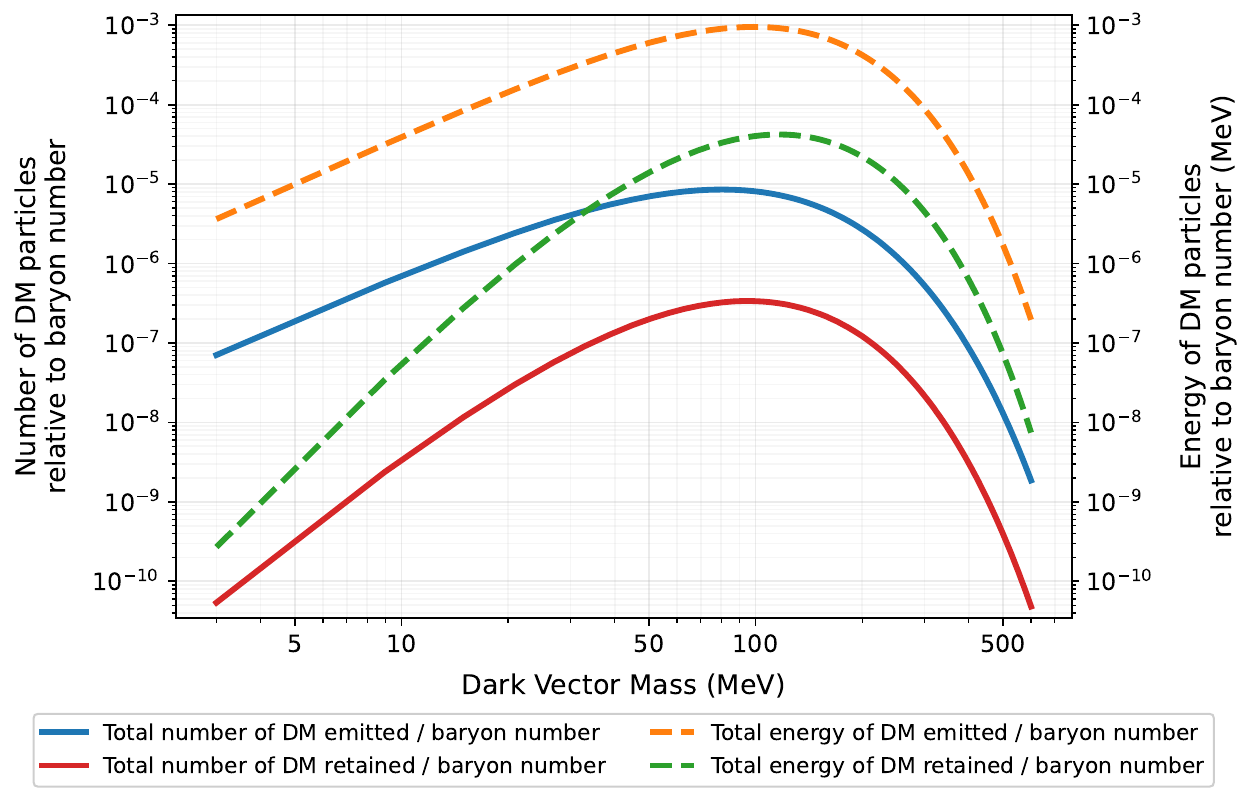}
\caption{Total number of DM particles produced and the DM energy in MeV per baryons (dashed curves). Realistic curves that take into account the retained fraction are shown in solid red (dashed green for DM energy retained per baryons). The coupling constant $\alpha_{\rm FI}$ is set to a value (\ref{a_FI_cosmo}) that reproduces observed DM abundance. }
\label{fig:Abundance}
\end{figure}

The total production of DM particles in SN is not a very useful quantity, as most of $\chi$ and $\bar\chi$ particles are too energetic to stay confined in the gravitational potential of the star. To that end, we evaluate the kinetic energy of DM particle required to escape the star, which varies from $\simeq 0.34 m_\chi$ at the center to $0.18m_\chi$ at the surface of the NS. We then calculate the fraction of produced DM that is confined to the star immediately after the SN explosion. This fraction depends rather sensitively on the $m_V/m_\chi$ ratio. For $m_\chi/m_V \to 1/2$ this fraction is close to 100\%, while for $m_\chi/m_V \to 0$, this fraction is minuscule. When $m_\chi/m_V \to 1/3$ as in this paper, the mechanism for producing a bound particle is the following: if $V$ decays at rest, both $\chi$ and $\bar\chi$ have enough velocity to leave NS. If on the other hand $V$ is boosted, then $\chi(\text{or}\ \bar\chi)$ emitted against velocity of $V$ can stay bound. Taking, for example $\Delta \Phi = \Phi_{r\to \infty}-\Phi(r)= 0.25$ (here $m_\chi\Phi(r)$ is the potential energy), one readily finds that the interval of boosts $\gamma_V$ that produces one particle confined to NS is from 1.036 to 2.71. Specifically, after an elementary calculation, one finds that the probability of retention ($m_V=3m_\chi$) is 
\begin{eqnarray}
\label{Pret}
    P^{ret}(\Delta \Phi,\gamma_V) = \frac12\frac{\Delta\Phi + 1 -\frac32\gamma_V +\frac{\sqrt{5}}{2}\sqrt{\gamma_V^2-1}}{\sqrt{5}\sqrt{\gamma_V^2-1}}.
\end{eqnarray}
Numerically it peaks at a few percent at $\gamma_V \sim 1.3$. 
Notice the coefficient $1/2$ in (\ref{Pret}) that comes from the fact that only one of the two particles produced can be retained by the star if $m_V/m_\chi=3$. 
This expression can be applied directly inside (\ref{simplified}) with the following substitution:
\begin{eqnarray}
    T(x,t)K_1\left(\frac{m_V}{T(x,t)}\right)\to ~~~~~\\\nonumber m_V\int_1^\infty P^{ret}(\Delta \Phi (x),\gamma_V)\sqrt{\gamma_V^2-1} \exp\left(-\frac{m_V\gamma_V}{T(x,t)}\right)d\gamma_V.
\end{eqnarray}
Determining Newtonian potential $\Delta \Phi (x)$ numerically, and evaluating these expressions as function of $m_\chi$, we plot the resulting abundance the energy stored per baryon in Fig.\,\ref{fig:Abundance}. 
This figure shows that, indeed, the FI production of DM inside the star is small but not negligible. In addition, we have checked the internal consistency of the calculations with different numerical inputs for $T(x,t)$, \cite{Roberts:2012zza} or \cite{Sumiyoshi:2022uoj}. Notice that for scenario A, we assume that the dark coupling constant gives a collision length larger than the size of the star, which in reality means $\alpha_d < 10^{-5}$. Above that value, one would have to consider a more complicated collective dynamics of $\chi$ fluid. As we shall see, the model is probed in the regime of very small self-interaction, $\alpha_d\sim O(10^{-10})$, and therefore collective effects during the FI process can be neglected. 

While gravitational retention for chosen values of parameters constitutes only a few percent from the total number of DM particles produced, it offers one interesting aspect that ensures the hierarchy of in- and out- reactions. Namely, the retained particles do not have enough kinetic energy to participate in the on-shell production of $V$, {\em i.e.} $\chi\bar\chi\to V$ is not possible.  This implies that reverse out-reaction has to proceed via an off-shell $V$ exchange and, therefore, is parametrically suppressed as $\sigma_{ann}\propto  g_d^2 g_L^2 \varepsilon^2$.

After the DM is produced and bound to the star by gravity, the entire system enters a long cooling regime, which is well researched in the proper NS literature \cite{Yakovlev:2004iq}. The main interest for us is the behavior of the DM gas bound to the NS by gravity. For scenario A, eventual DM-DM scattering leads to thermalization and the establishment of some ``dark" temperature $T_d$. This temperature is below the average escape energy, as every particle initially is below this threshold. However, with time, a small Boltzmann tail of the distribution can evaporate away. This evaporation leads to the reduction of $T_d$ and $N_\chi$. The cooling occurs faster than the particle loss, due to the fact that only the highest energy particles can evaporate,
\begin{eqnarray}
    \frac{d\log N_\chi(t)/dt }{d\log T_d(t)/dt }\ll 1. 
\end{eqnarray}
We estimate that the loss of particles as a result of thermalization and evaporation is relatively small. In scenario B, the self-interaction is negligible (as $\alpha_d = \alpha_{\rm FI} \sim O(10^{-26})$), but the thermalization may occur on a very long time scale due to scattering on muons ($\mu^-$) present at a percent level relative to other particles in the interior of NS. This cooling process is more uncertain but may occur on time scales comparable with annihilation. 

Finally, the reverse process of Fig.\,\ref{fig:F_diagrams}, the annihilation of the DM particles, occurs on very long time scales due to the combination of the following factors. Firstly, the rate scales as $n_\chi^2$, and since the DM is dilute, this naturally elongates the time scale involved. Secondly, as already mentioned, the annihilation rate is multiplied by additional small coupling constants as it is mediated by an off-shell $V$. The annihilation of DM injects energy via
\begin{eqnarray}
    \chi\bar\chi\to \nu_{\mu(\tau)}\bar\nu_{\mu(\tau)};~ \nu_i + N(e) \to \nu_i+ N(e).
    \label{injection}
\end{eqnarray}
The products of DM annihilation, muon and tau neutrinos, undergo scattering on nucleons and electrons mediated by the $Z$ exchange. For relevant energies $E_\nu = m_\chi$, the collision length is short, $(\langle \sigma_{\nu N} \rangle n_N)^{-1}\ll R_{\rm NS}$, due to extremely dense NS interior.  This makes the neutrinos to diffuse via a process of multiple collisions and deposit $O(1)$ fraction of their initial energy inside NS. Thus, the rate of the energy injection is determined by the first reaction in (\ref{injection}), for which  we need to know the annihilation cross section and the distribution of $\chi$ particles over the radius, as well as the change of total DM particle number in time, $N_\chi(t)$. The relevant set of equations is given by:
\begin{eqnarray}
\frac{dQ}{dt}=-2m_\chi\frac{dN_\chi}{dt}=-\frac{2m_\chi \langle \sigma_{ann}v\rangle N_\chi N_{\bar \chi}}{V_{\rm NS}}\times J,~~\\\nonumber
J\equiv V_{\rm NS} \int \frac{n_\chi(r) n_{\bar\chi}(r)}{N_\chi N_{\bar\chi}}dV= V_{\rm NS}\int {f_\chi(r) f_{\bar\chi}(r)}dV.
\end{eqnarray}
In these expressions, the dimensionless factor $J$ parametrizes deviation from a uniform distribution. Gravity over-concentrates DM particles in the NS center, and $J>1$. In the limit of uniform distribution $f_\chi\to V_{\rm NS}^{-1}$, $J\to1$. Small amount of kinetic energy carried by $\chi$ particles is neglected. Note that the annihilation cross section is taken outside of the volume integral, which is the approximation held due to $s$-wave, velocity-independent, nature of $\sigma_{ann}v$ in our model. Specifically, for scenario A, we have:
\begin{eqnarray}
    \sigma_{ann}v = \frac{\pi\alpha_{\rm FI}\alpha_d }{m_\chi^2}\, G\left(\frac{m_V}{m_\chi},\frac{m_X}{m_\chi}\right)\to 0.64\times\frac{\pi\alpha_{\rm FI}\alpha_d }{m_\chi^2},~~~\\\nonumber
    G=\frac{(4m_\chi^2)^4(m_V^2-m_X^2)^2}{m_V^4(4m_\chi^2-m_V^2)^2(4m_\chi^2-m_X^2)^2};~G(3,0) = 0.64 
    \label{sigma_ann}
\end{eqnarray}
where $v$ is the relative velocity of two DM particles. A rather cumbersome ratio of propagators, denoted by $G$, arises because $\chi\bar\chi$ production occurs at $s=m_V^2$ but annihilation occurs at $s=4m_\chi^2$. For scenario B, this factor will be slightly modified, and $\varepsilon^2\alpha_L$ will take place of $\alpha_d$ in Eq.\,(\ref{sigma_ann}).

\begin{figure}[t]
\includegraphics[width=0.42\textwidth]{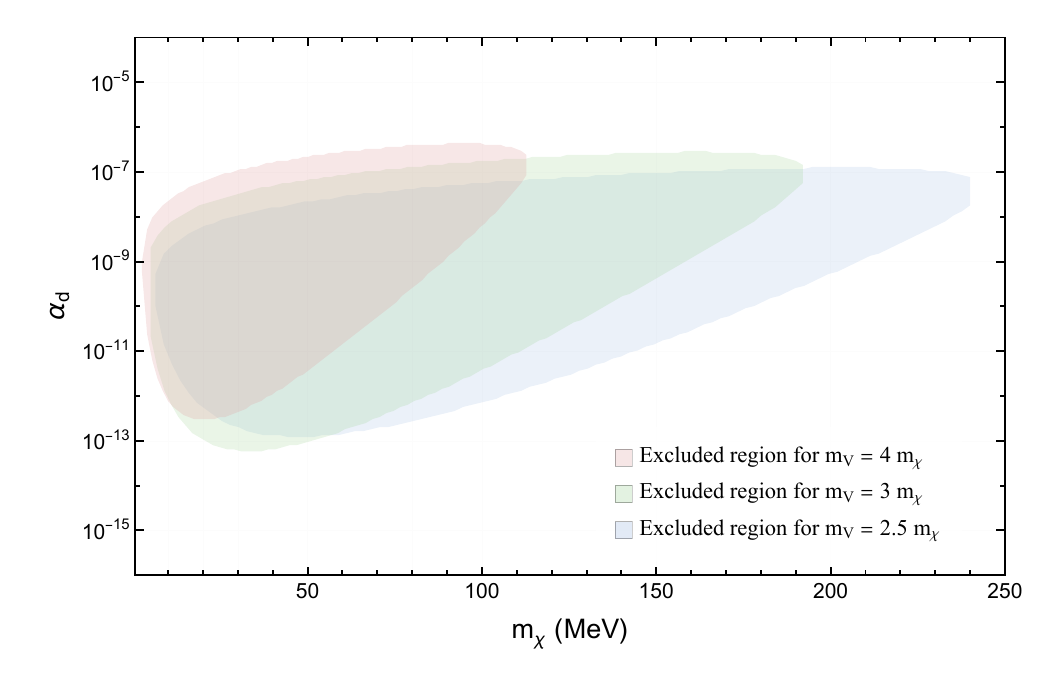}\\
\includegraphics[width=0.42\textwidth]{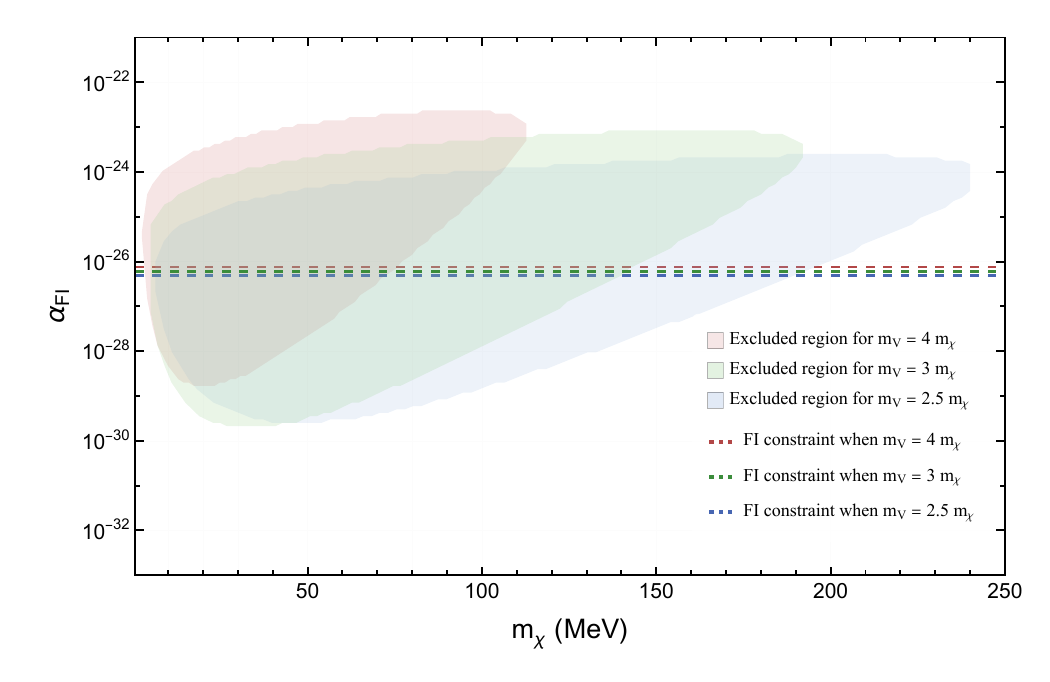}
\caption{NS heating constraints as exclusion plots for certain projections of the parameter space. Top panel: NS heating constraints on $\{m_\chi,\alpha_d\}$ parameter space when $\alpha_{\rm FI}$ is fixed to reproduce observed amount of DM, and $m_V/m_\chi$ is varied for three representative values. Bottom panel: dark coupling constant is fixed, $\alpha_d =10^{-10}$, and constraints are shown on $\{m_\chi,\alpha_{\rm FI}\}$ plane. Dashed lines represent ``optimal value" consistent with $\chi$ saturating DM abundance.  }
\label{fig:final_results}
\end{figure}

As already mentioned, we utilize observational limits on abnormal sources of energy for $\text{PSR J2144}- \text{3933}$ \cite{Guillot:2019ugf},
\begin{eqnarray}
    \frac{dQ}{dt}(t\sim 10^{16}\,{\rm s}) \lesssim 2.3\times 10^{39}\,\rm\frac{eV}{s}.
\end{eqnarray}
Returning to the schematic Fig.\,1 in this paper, we note that the coupling constant $\alpha_d$ plays the role of a ``dial" for $R_{out}/R_{in}$. Too large value for $\alpha_d$ leads to premature/early annihilation, leaving the late time abundance $N_\chi$ suppressed by $\sim \sigma_{ann}^{-1}$, while too small value does not lead to sufficient energy injection at late times. We are now prepared to make a final synthesis of all the ingredients and derive constraints on the parameter space of the model. We evaluate it for scenario A, taking $T_d$ to be small, and finding the radial equilibrium distribution $f_\chi(r)$. Enhancement factor $J$ typically stays in the range of $O(10)$. 

Final results are plotted in Fig.\,\ref{fig:final_results} for scenario A (very similar figure for scenario B with $\alpha_d \to \varepsilon^2\alpha_L$ substitution.) The figure shows that, as expected, constraints apply to DM mass  from 10\,MeV to about 100\,MeV range. For $\alpha_d > 10^{-6}$ the annihilation occurs early, and the NS heating constraints do not constrain the model.  Dark coupling constant $\alpha_d \sim 10^{-10}$ produces enough suppression to delay $\chi\bar\chi\to \rm SM$ annihilation. As bottom panel shows, NS heating excludes a range of cosmologically viable models that are not possible to constrain in any other way. At large mass, there is an exponential cutoff of the produced $V$ particles. At low mass, the produced vector particles are relativistic, and the fraction of their daughter $\chi$ particles retained by the star is very small. We have also varied $m_V/m_\chi$ to show that, as expected, higher values of the ratio result in smaller retention probability $P^{ret}$ and weaker constraints.

Finally, it is instructive to estimate the scattering cross-section of $\chi$ particles on electrons to show that direct detection experiments are not competitive in this class of models. While atoms do not contain particles from the second and third generation, the kinetic mixing between photons and $X$ mediator is inevitable. Taking this parameter to be $\epsilon_{AX} \sim 10^{-2} \times g_L e$, we arrive at the following estimate of the $\chi$-electron scattering cross section: 
\begin{eqnarray}
    \sigma_{\chi e} \sim 16\pi (e\epsilon_{AX}/g_L)^2 \alpha_d\alpha_{\rm FI}\times \frac{m_e^2}{m_V^4} \propto 10^{-70} \,{\rm cm}^2,
\end{eqnarray}
where we took $\alpha_d \sim 10^{-10}$, and $m_V=100$\,MeV. This cross section is more than twenty five orders of magnitude below current detection capabilities.  

{\bf Conclusions.}
Late time heating of NS from the freeze-in dark matter is a new powerful way of constraining stable DM particles in the MeV-to-100\,MeV range. Thermal production of DM during the SN explosion does not affect energetics of the explosion as only $O(10^{-6})$ particles per nucleon are produced. Yet the energy stored in these particles will be returned back to the star via the annihilation process that can be very slow compared to the pair creation. Existing data put constraints on models where $R_{out}\ll R_{in}$ as demonstrated in this paper on the example of the extremely weak leptonic-force-mediated FI dark matter model. Our results motivate investigations of other well-motivated FI models, including those with dark photon and scalar Higgs portal mediation. It also motivates investigations of the FI-DM-induced collapse of NS into a black hole for scalar DM (see {\em e.g.} Refs. \cite{Bramante:2023djs,Kouvaris:2011fi}), and electromagnetic emission signatures from the FI DM annihilating outside the NS radius.

 {\bf Acknowledgments.} 
We would like to thank Drs. J. Bramante,  G. Raffelt,  S. Reddy, B. Safdi and E. Vitagliano for valuable conversations. 
The authors are supported in part by U.S. Department of Energy Grant No. DE-SC0011842.

\bibliography{ref}
\end{document}